\documentclass[12pt]{article}
\usepackage{lineno,hyperref}
\usepackage[section]{placeins}
\usepackage[table]{xcolor}
\usepackage{lipsum}
\usepackage{multirow}
\usepackage{setspace}
\usepackage{anysize}
\usepackage{graphicx}
\usepackage{epstopdf}
\usepackage{color}
\usepackage{url}
\usepackage{latexsym}
\usepackage{amsmath}
\usepackage{amssymb}
\usepackage{amsfonts}
\usepackage{syntonly}
\usepackage{titlesec}
\usepackage[round]{natbib}

\newcommand{\wv}{\psi(t)}
\newcommand{\sca}{\phi(t)}
\newcommand{\be}{\begin{eqnarray}}
\newcommand{\ee}{\end{eqnarray}}
\newcommand{\ba}{\begin{eqnarray*}}
\newcommand{\ea}{\end{eqnarray*}}
\newcommand{\R}{\mathbb{R}}
\newcommand{\bt}{\mathbf{t}}
\newcommand{\bs}{\mathbf{s}}
\newcommand{\EE}{\mathbb{E\,}}
\newcommand{\var}{\mathbb{V}{\bf{ar\,}}}
\newcommand{\ZZ}{\mathbb{Z}}

\begin{document}

\noindent ROBUST WAVELET-BASED ASSESSMENT OF SCALING WITH APPLICATIONS
\vskip 3mm

\vskip 5mm
\noindent Erin K. Hamilton$^{\rm a}$,  Seonghye Jeon$^{\rm a}$, \\
Pepa Ram\'irez Cobo$^{\rm b}$, Kichun Sky Lee$^{\rm c}$ and Brani Vidakovic$^{\rm d}$

\noindent $^{\rm a}$  Georgia Institute of Technology, Atlanta, GA, US

\noindent $^{\rm b}$  Universidad de C\'adiz, C\'adiz, Spain

\noindent $^{\rm c}$  Hanyang University, Seoul, Korea

\noindent $^{\rm d}$  Texas A\&M University, College Station, TX, US

\vskip 3mm

\noindent ABSTRACT

 A number of approaches have dealt with statistical assessment of self-similary, and many of those are based on multiscale concepts. Most rely on certain distributional assumptions which are usually violated by real data traces, often characterized by large temporal or spatial mean level shifts, missing values or extreme observations. A novel, robust approach based on Theil-type weighted regression is proposed for estimating self-similarity in two-dimensional data (images). The method is compared to two traditional estimation techniques that use wavelet decompositions; {ordinary least squares} (OLS) and Abry-Veitch  bias correcting estimator (AV). As an application, the suitability of  {the self-similarity estimate resulting from the} the robust approach is illustrated {as a predictive feature} in the classification of digitized mammogram images as cancerous or non-cancerous. The diagnostic employed here is based on the properties of image backgrounds, which is typically an unused modality in breast cancer screening. Classification results show nearly 68\% accuracy, varying slightly with the choice of wavelet basis, and the range of multiresolution levels used.
%

\newpage

\noindent 1.   INTRODUCTION

{High-frequency} signals and high-resolution digital images common in real-life settings often possess a noise-like appearance.
Examples of such signals have been found in a variety of systems and processes including economics, telecommunications, physics, geosciences, as well as in biology and medicine \citep{Engel2009, Gregoriou2009, Katul2001, Park2000,   Woods2016, Zhou1996}. Often, statistical descriptions of noise-like signals and images involve the degree of their irregularity as a key statistical summary. Conditional on appropriate stochastic structure of the signals, irregularity measures can be tied with measures of self-similarity, fractality, and long memory.

{High-frequency} signals, whether naturally occurring or human-generated, usually show substantial \emph{self-similarity}.
 Formally, a deterministic function $f(t)$ is said to be
self-similar  if $f(t)=a^{-H}f(a t)$, for some choice of the exponent
$H$, and for positive dilation factors $a$.  The notion of
self-similarity has been extended to random processes where
the equality of functions is substituted by an equality in
distribution of random variables. Specifically, a
stochastic process $\{X(t),\ t\in \R\}$ is self-similar with
scaling exponent (or \emph{Hurst exponent}) $H$ if, for any
$a \in \R^+$,
\begin{equation}\label{basic def}
X(a t)\overset{d}{=}a^H X(t),
\end{equation}
{where $\overset{d}{=}$ denotes equality of all joint
finite-dimensional distributions.}

Many methods (either defined in time or scale/frequency domains) for
estimating $H$ in one dimension exist. For a comprehensive
description, see \cite{Beran1994}. In particular, the discrete and
continuous wavelet transforms \citep{Daubechies1992, Mallat1998} have proven
suitable for modeling self-similar processes
with stationary increments, as the fractional Brownian
motion (fBm) \citep{Abry2000, Abry2001, Abry2003}.
Wavelet-based methods for estimating $H$ have
been proposed in literature for the 1-D case \citep{Audit2002, Soltani2004, Veitch1999}.
However, none of these methods take into account violations in model assumptions usually presented by
real data sets. In particular, several real-life sources involve systematic frequency-dependent noise which
induces non-Gaussianity in the time domain, and consequently in the wavelet domain as well. The presence of outlier multiresolution levels, inter and between level dependencies and distributional contaminations make
the robust estimation of $H$ an issue of interest.
Some robust approaches for estimating self-similarty have been recently examined in literature \citep{Franzke2012, Park2009, Shen2007, Sheng2011}.

In this paper, a robust approach in estimating $H$ in self-similar signals is considered.
Here the focus is on images as the selected application, but the methodology applies to a multiscale
context of arbitrary dimension in which a hierarchy of multiresolution subspaces can be identified as a generator of spectra. The approach is based on a Theil-type weighted regression \citep{Theil1950} where
average multiresolution level ``energies,'' that is, squared wavelet coefficients, are regressed against the level indices. The performance
of the robust approach is compared with two benchmark approaches: ordinary least squares (OLS) and Abry-Veitch (AV) method. See \cite{Katul2001} and \cite{Veitch1999}, respectively. 

As an application, the suitability of the proposed estimator  {as a predictive feature} in classification of digitized mammogram images as cancerous or non-cancerous is demonstrated. Many medical images possess scaling characteristics that are discriminatory. The proposed Theil-type estimator is applied as a possible  {predictive measure for inclusion in screening technologies}. Most of the references found in literature dealing with automated breast cancer detection {in mammography} are based on microcalcifications \citep{El-Naqa2002, Kestener2001, Kiran2014, Netsch1999, Wang1998}.
A comparative overview of machine learning approaches in BC diagnostic can be found in \cite{Kourou2015}.
Only recently has scaling information found in background tissue come into consideration \citep{Erin2011, Nicolis2011, Pepa2013, Jeon2015, Roberts2017}.
 {For this predictive measure}, the focus is on the scaling information from the entire image rather than localized features traditionally used. Adding the proposed method to an existing battery of established tests has a potential to improve the overall accuracy of mammogram screening techniques.

This paper is organized as follows. Section 2 gives background on 2-D discrete wavelet transforms with a review of wavelet-based spectrum in the context of estimating $H$ for fractional Brownian motion. Section 3 is devoted to statistical estimation of $H$. In Section 3.1 the benchmark non-robust approaches for comparison are described. In Section 3.2 our robust approach is presented, with Section 3.3 illustrating the performance of the new technique on simulated data sets. In Section 4 the performance of our robust approach in differentiating between cancerous versus non-cancerous tissue in mammogram images is assessed. Finally, this paper is concluded with remarks and recommendations for practical use of the methodology. Technical details concerning the newly introduced robust measure discussed in Section 3 are deferred to Appendix A. Appendix B contains more extensive simulations, and for space considerations is available online.

\vskip 4mm

\noindent 2. BACKGROUND


\vskip 4mm

\noindent 2.1 The 2-D Discrete Wavelet Transform

A review of the 2-D discrete wavelet transform builds upon the 1-D orthogonal wavelet decomposition, which can express any square integrable function $X\in \mathcal{L}_2(\mathbb{R})$ in terms of shifted and dilated versions of a wavelet function $\wv$ and
shifted and dilated versions of a scaling
 function $\sca$. A detailed introduction to wavelet theory can be found in classic monographs by \cite{Daubechies1992} or \cite{Mallat1998}. Many
signals arising in practical applications are multidimensional, including our current application of mammogram images. The 1-D wavelet transform is readily generalized to the multidimensional case.

The 2-D wavelet basis atoms are
constructed via translations and dilations of a tensor product of
univariate wavelet and scaling functions:
\begin{eqnarray} \nonumber
\phi(t_1,t_2)&=&\phi(t_1)\phi(t_2),\\  \nonumber
\psi^h(t_1,t_2)&=&\phi(t_1)\psi(t_2),\\ \nonumber
\psi^v(t_1,t_2)&=&\psi(t_1)\phi(t_2) \; ~\text{and} \\ \label{psid}
\psi^d(t_1,t_2)&=&\psi(t_1)\psi(t_2).
\end{eqnarray}
The symbols $h,v,d$ in (\ref{psid}) stand for horizontal, vertical
and diagonal directions, respectively. Consider the wavelet atoms,
\begin{eqnarray}\label{atom1}
\phi_{j,\mathbf{k}}(\mathbf{t})&=&2^{j}\ \phi(2^jt_1-k_1, 2^jt_2-k_2) \; ~\text{and} \\ \label{atom2}
\psi^i_{j,\mathbf{k}}(\mathbf{t})&=&2^{j}\ \psi^i(2^jt_1-k_1, 2^jt_2-k_2),
\end{eqnarray}
\noindent for $i\in \{h,v,d\}$, $j\in \mathbb{Z}$, $\mathbf{t}=(t_1,t_2)\in
\mathbb{R}^2$, and $\mathbf{k}=(k_1,k_2)\in \mathbb{Z}^2$. Then, any function $X\in \mathcal{L}_2(\mathbb{R}^2)$ (an image, for example) can be
represented as
\begin{equation}\label{wav_f}
X(\mathbf{t})= \sum_{\mathbf{k}} c_{J_0\mathbf{k}}\phi_{J_0,\mathbf{k}}(\mathbf{t})+ \sum_{i \in \{h,v,d\}} \sum_{j\geq J_0}\sum_{\mathbf{k}}  d^i_{j,\mathbf{k}}\psi^i_{j,\mathbf{k}}(\mathbf{t}),
\end{equation}
\noindent where the wavelet coefficients are given by

\begin{equation*}
d^i_{j,\bf{k}}=2^{j}\int X(\mathbf{t})\ \psi^i (2^j \mathbf{t} - \mathbf{k}) \;d\bf{t},
\end{equation*}
and $\mathcal{L}_2(\mathbb{R}^2)$ is the space of all real square integrable 2-D functions. In expression (\ref{wav_f}), $J_0$ indicates the coarsest scale or lowest resolution level of the transform, and larger $j$ correspond to higher resolutions.

\vskip 4mm

\noindent 2.2 The 2-D fBm: Wavelet Coefficients and Spectra

Consider a self-similar stochastic process $\{X(t),\ t\in \R\}$ as in (\ref{basic def}). Then, detail  coefficients defined satisfy
 $$\displaystyle d_{jk}\overset{d}{=}2^{-j(H+1/2)}d_{0,k},$$
 for a fixed level $j$ and under $\mathcal{L}_2$ normalization \citep{Abry2001, Flandrin1992}. If, in addition, the process has stationary increments (i.e., $X(t+h) - X(t)$ is independent of $t$), then $\mathbb{E}\left( d_{0k}\right)=0$ and
$\mathbb{E}\left( d^2_{0k}\right)=\mathbb{E}\left( d^2_{00}\right)$. Therefore,
 \begin{equation}\label{fundamental relation}
 \mathbb{E}\left(d^2_{jk}\right)\propto 2^{-j(2H+1)},\end{equation}
 which provides a basis for estimating $H$ by taking logarithms on both sides of equation (\ref{fundamental relation}). The sequence, $S(j)=\log \mathbb{E}\left(d^2_{jk}\right)$, where $j\in \mathbb{Z}$,   is called the  wavelet spectrum. \cite{Veitch1999} explored in detail   wavelet spectra and statistical estimation of the $H$ under the assumption that the process $X(t)$ is Gaussian.
 When the Gaussianity is combined with $H$-self-similarity, as in (\ref{basic def}), and independent increments, the resulting stochastic process is unique.
  It is called fractional Brownian motion (fBm) and denoted as $B_H(t)$.  This process is arguably  the most popular model for signals that scale.

The definition of the one-dimensional fBm can be readily extended to the multivariate case \citep{Levy1948}, and more recently, to the case of vector fields. A two-dimensional fBm, $B_H(\bt)$, for $\bt$ $\in
[0,1]\times[0,1]$ and $H \in (0,1)$, is a Gaussian process with stationary
zero-mean increments, for which (\ref{basic def}) becomes
\begin{equation*}
B_H(a{\mathbf t})\overset{d}{=}a^H B_H(\bt).
\end{equation*}
The auto-covariance function is given by
\begin{equation}\label{cov}
\mathbb{E}\left[B_H(\bt)B_\emph{H}(\bs)\right]=\frac{\sigma_{H}^2}{2}
\left(\|\bt\|^{2\emph{H}}+\|\bs\|^{2\emph{H}}-\|\bt-\bs\|^{2\emph{H}}\right),\end{equation}
where $\sigma_{H}^2$ is a positive constant depending on $H$, and $\|\cdot\|$ is the usual Euclidean norm in $\mathbb{R}^2$. Because of the specific structure (\ref{cov}), it can be shown \citep{Flandrin1992, Reed1995} that the expected values of the detail coefficients associated to the 2-D fBm
satisfy
\begin{equation}
E\left[ \left\vert d_{j,\mathbf{k}}^{i}\right\vert ^{2}\right] =\frac{%
\sigma^2_H}{2} V_{\psi ^{i}}2^{-(2H+2)j},  \label{varfinal0}
\end{equation}
where
$V_{\psi ^{i}}$ depends only on the wavelet $\psi^i$ and exponent $H$, but not on the scale $j$. Equivalently,
\begin{equation}
\log_{2}\mathbb{E}\left[ \left\vert d_{j,\mathbf{k}}^{i}\right\vert
^{2}\right] =-(2H+2)j+C_{i}, \label{spectra2}
\end{equation}
which defines the two-dimensional wavelet spectrum $S^i(j)$, from which $H$ can be estimated.  The next section will consider statistical estimation of $H$ in a 2-D fBm, from this spectrum.

\vskip 4mm

\noindent 3. Statistical Estimation in 2-D fBm

The form of wavelet spectra and the relationship between $H$ and scale index $j$ provide a natural
way of estimating scaling. An overview is given of two benchmark approaches based on spectral regression
that are typically used in practice -- ordinary least squares (OLS) and the Abry-Veitch (AV) method. The proposed Theil-Type (TT) estimator is then introduced.

\vskip 4mm

\noindent 3.1 Two Benchmark Approaches

Equation (\ref{spectra2}) points toward a linear regression procedure to estimate $H$ from the slope of the regression
when $\log_{2}\mathbb{E}\left[ \left\vert d_{j,\mathbf{k}}^{i}\right\vert
^{2}\right]$ is regressed on the level $j$.
 The first traditional estimate obtained using such an approach is ordinary least squares (OLS). As detailed in \cite{Veitch1999}, two main complications arise when considering the linear regression in (\ref{spectra2}). The first is that $\mathbb{E}\left[ \left\vert d_{j,\mathbf{k}}^{i}\right\vert
^{2}\right]$ is not known but must be estimated. However, the near-decorrelation property \citep{Abry2003, Craigmile2005} of the wavelet coefficients (which also holds for the 2-D coefficients, $d_{j,\mathbf{k}}^{i}$) validates the use of the empirical counterpart
$$
\mu_j^i = \frac{1}{n_j}\sum_{\bf k} |d^i_{j,\bf k}|^2,
$$
where the summation is
made over all two-dimensional shifts $\bf k$ within the multiresolition level $j$ from the hierarchy $i$,
and $n_j$ denotes the total number of coefficients at that level.
For example, for a square dyadic-side image, $n_j = 2^{2 j}.$

The OLS regression defined on pairs
\begin{eqnarray}
\left(j, \log_{2}\mu_j^i\right), \quad i=h, v,d,
\label{empspectra2}
\end{eqnarray}
is typically used as a computationally inexpensive method which, for some cases, has proven to work well in practice. See, for example, \cite{Nicolis2011}. Thus it is used by many for first-attempted estimations. According to this method, $\hat{H}=-(s + 2)/2$, where $s$ denotes the slope of the regression. Although OLS ignores
the fact that regression leading to estimation of $H$ is heteroscedastic, our experience is that when the length of a signal is large, the corrections for heteroscedasticity, dependence, and bias are reasonably small compared to inherent noise in the simulations or real data.

As mentioned, the assumption of homoscedascticity of errors, tacitly assumed for OLS, is
violated.
In addition, the logarithm  for base 2 of $\mu_j^i,$  taken as an estimator of $\log_2 \EE(d^2),$ is biased.

According to \cite{Veitch1999},
$$
\var \left(\log_{2}\mu_j^i\right) \sim \frac{2}{n_j \log^2 2}.
$$
Since the variances vary with the level $j$, a weighted regression
is thus more adequate in the context. In \cite{Veitch1999}, a bias correction term is proposed as well,
by replacing
$\log_{2}\mu_j^i $ in
(\ref{empspectra2}) by
$\log_{2}\mu_j^i + 1/(n_j \log 2).$

Thus, the second traditional estimate of the $H$, the AV estimate, is obtained from the slope of bias-corrected weighted linear regression with weights given by
$$
w_{j} \propto \frac{n_j \log^2 2}{2}.\quad
$$

Although AV accounts for the differences in variances at each level, this method still assumes that the errors are normally distributed at each level. In fact, $\log_{2}\mu_j^i$ is distributed as the logarithm of a chi-squared variable, which is non-symmetric about its location.

\vskip 4mm

\noindent 3.2 Proposed Theil-type Estimator

Real-world signals (as network traffic traces) may be characterized by non-stationary conditions such as sudden level shifts, breaks or extreme values; see for example \cite{Shen2007}. The outlier levels in the observed data, often caused by the instrumentation noise, would leave a bump or a ``hockey stick'' signature in the wavelet spectra,
thus violating the conditions assumed for theoretical benchmark processes, such as fBm. Therefore, it is desirable to employ robust approaches while estimating
scaling indices. Recently, there has been an interest in such an approach \citep{Franzke2012, Park2009, Shen2007, Sheng2011}.
These works focus on the estimation of self-similar signals in one dimension and adopt different approaches than the methodology proposed here to achieve robustness.

In the rest of this section, a technique is introduced for robust estimation of $H$ for
two dimensional signals (images) and its theoretical properties are derived
on 2-D fBm, as a calibrating process. The approach is based on the Theil-type estimator,
a method for robust linear regression that selects the weighted average
of all slopes defined by different pairs of regression points \citep{Theil1950}.
This estimator is less sensitive to outlier levels and
can be significantly more accurate than simple linear regression for skewed and heteroskedastic data. The main benefit is the case when the processes are not exactly monofractal  but contain outlier levels
that affect the linearity of the spectra, especially at coarse levels. On the other hand, this method is comparable to non-robust regression methods for normally distributed data in terms of statistical power \citep{Wilcox2001}.

In this paper, a weighting scheme is adopted under which each pairwise slope is weighted by an inverse of the variance of the estimated slope for that pair, as in \cite{Birkes1993}, \cite{Jaeckel1972},
and \cite{Sievers1978}. Specifically, the slopes of the linear equations in (\ref{spectra2}) are assessed as a weighted average of all pairwise slopes between levels $i$ and $j$, $\{s_{ij}\}$, with weights satisfying
\ba
w_{ij} \propto \left(i-j\right)^2 \times HA\left(2^{2i}, 2^{2j}\right),
\ea
where $HA$ is the harmonic average. Thus the proposed estimator is robust with respect to possible outlier levels and free of any distributional assumptions. As seen in Appendix A, which contains their full derivation, weights for each pair are designed to reduce the undue influence that outliers can have on estimates. Specifically, the influence of the coarse levels that in reality show more instability is additionally de-emphasized by weighting choices. Finally, the estimator of the overall slope (by which the parameter $H$ is estimated) is given by
\ba
\sum_{i<j} w_{ij} s^*_{ij} / \sum_{i<j} w_{ij},
\ea
where
\ba
s_{ij}^* = s_{ij} + \frac{1}{(j-i)\log 2} \left( \frac{1}{2^{2j}} - \frac{1}{2^{2i}} \right)
\ea
is the bias-corrected pairwise slope between $i$th and $j$th points. This new estimation approach will be denoted as TT, short for Theil-type.

\vspace*{0.1in}

\noindent
{\bf Remark.~} Although for the case of 2-D fBm, the slope $s$ and consequently
the Hurst exponent $H=-(s + 2)/2$,   theoretically coincide
for all three hieararchies of multiresolution spaces
$\{d, h, v\},$
in practice we obtain the estimators of three slopes $s_i,$ $i \in \{d, h, v\}.$
Consequently, there would be three estimators of $H$,
 $\hat{H}_i =-(s_i + 2)/2$.

From extensive simulations for isotropic fields, it is concluded that the estimator
$\hat{H}_d$ obtained from diagonal hierarchy often suffices in estimating $H$,
and that estimators  $\hat{H}_h$ and   $\hat{H}_v$ bring little new information.
This agrees with findings in \cite{Nicolis2011}, \cite{Pepa2013}, and \cite{Jeon2015}.

\vskip 4mm

\noindent 3.3 Simulations and Comparisons

To illustrate the performance of the robust method described in the previous section, consider the next simulation example. A total of $100$ realizations of one-dimensional fBm of length 512 and $100$ realizations of 2-D fBm size 512 $\times$ 512, each characterized by Hurst exponents $H \in \{0,3,\ 0.4,\ 0.5,\ 0.6,\ 0.7\},$ were simulated. The one-dimensional fractitional Brownian motion was simulated based on the method of \cite{Wood1994} and \cite{coeurjolly2000simulation}, and the two-dimensional fractional Brownian motion was simulated using Barri\'ere's Matlab code \citep{Barriere2007}. The code can be found at \url{http://gtwavelet.bme.gatech.edu/}.

  A wavelet transform was then performed on the simulated data, using  Haar, Coiflet 4 tap, Daubechies 6 tap, and Symmlet 8 tap wavelet filters. The estimated Hurst exponents were obtained using the two standard methods described, OLS and AV, and the robust method, TT.

To mimic realistic data that in their wavelet decompositions often show instability at coarse
 levels of detail, the procedure is repeated with the same realisations but contaminated at a coarse level. This is done by adding white noise of zero mean and variance $\sigma^2_{ij}$, where $\sigma^2_{ij}$ is the average variance of wavelet coefficients at direction $i \in \{d, h, v\} $ and level $j$. For this simulation, wavelet coefficients were contaminated at level 3 and wavelet spectra was calculated from levels 3 through 7.

Tables \ref{tab:1d_original} and \ref{tab:1d_contaminated} report the estimated values of $H$ for $H=0.5$, for the non-contaminated and contaminated 1-D cases. Also the mean squared errors (MSE), as a sum of both the bias-squared and the variance of the estimates, are provided. Cells with underlined values represent lowest bias, and the grayed cells indicate the cases with lowest MSE.

\begin{table}[!h]
\begin{center}
\caption{Estimations of $H$ and MSEs for $H=0.5$ under four different wavelet filters, in the non-contaminated case.}
\vspace{.2in}
\scalebox{0.9}{\begin{tabular}{c  l  c >{\centering}p{1.7cm} c >{\centering}p{1.7cm} c >{\centering}p{1.7cm} c c} \hline
\multicolumn{2}{c}{} && Haar && Coiflet4 && Daub6 && Symmlet8 \\  \hline
\multirow{2}{*}{OLS} & $H$ && \textbf{0.434} && \underline{\textbf{0.455}} && \textbf{0.460} && \textbf{0.456} \\
& \textit{MSE} &&0.011 && 0.009 && 0.010 && 0.010 \\ \hline
\multirow{2}{*}{AV}& $H$ &&\textbf{0.424} && \textbf{0.401} && \textbf{0.446} && \textbf{0.425} \\
& \textit{MSE} && 0.011&& 0.015 && 0.007 && 0.010 \\ \hline
\multirow{2}{*}{TT} & $H$ && \underline{\textbf{0.454}} && \textbf{0.446} && \underline{\textbf{0.479}} && \underline{\textbf{0.462}}\\
&\textit{MSE} && \cellcolor{black!20}0.007 && \cellcolor{black!20}0.008 && \cellcolor{black!20}0.005 && \cellcolor{black!20}0.006 \\
\hline
\end{tabular}}
\label{tab:1d_original}
\end{center}
\end{table}

\begin{table}[!h]
\begin{center}
\caption{Estimations of $H$ and MSEs for $H=0.5$ under four different wavelet filters, in the contaminated case.}
\vspace{.2in}
\scalebox{0.9}{\begin{tabular}{c  l  c >{\centering}p{1.7cm} c >{\centering}p{1.7cm} c >{\centering}p{1.7cm} c c} \hline
\multicolumn{2}{c}{} && Haar && Coiflet4 && Daub6 && Symmlet8 \\  \hline
\multirow{2}{*}{OLS} & $H$ && \textbf{0.535} && \textbf{0.548} && \textbf{0.541} && \textbf{0.552}\\
& \textit{MSE} && 0.014 && 0.014 && 0.015 && 0.017\\ \hline
\multirow{2}{*}{AV} & $H$ && \textbf{0.469} && \textbf{0.470} && \underline{\textbf{0.481}} && \textbf{0.472}\\
& \textit{MSE} && \cellcolor{black!20}0.007 && \cellcolor{black!20}0.007 && \cellcolor{black!20}0.007 && \cellcolor{black!20}0.006\\ \hline
\multirow{2}{*}{TT} & $H$ && \underline{\textbf{0.516}} && \underline{\textbf{0.520}} && \textbf{0.522} && \underline{\textbf{0.523}}\\
& \textit{MSE} && 0.008 && 0.007 && 0.008 && 0.008\\
\hline
\end{tabular}}
\label{tab:1d_contaminated}
\end{center}
\end{table}

From Table \ref{tab:1d_original}, it can be seen that TT estimates show the best performance with respect to both MSE and bias alone. In the case of Table \ref{tab:1d_contaminated}, where results have been obtained under a contamination in the original realizations, it can be deduced that AV and TT perform comparably well.

Tables \ref{tab:2d_original} and \ref{tab:2d_contaminated} summarize the estimated $H$ and MSE for 2-D realizations, in non-contaminated and contaminated scenarios, respectively.
Table \ref {tab:2d_original} shows that the OLS performs best, followed by TT, with respect to both MSE and bias. When the traces are contaminated, TT outperforms both OLS and AV in most settings, as shown in Table \ref{tab:2d_contaminated}.

\begin{table}[!h]
\begin{center}
\caption{Estimations of $H$ and MSEs for $H=0.5$ from three directions; under four different wavelet filters, in the non-contaminated case.}
\scalebox{0.9}{\begin{tabular}{c  l  c>{\centering}p{1.7cm}c>{\centering}p{1.7cm}c>{\centering}p{1.7cm}c  c>{\centering}p{1.7cm}c>{\centering}p{1.7cm}c>{\centering}p{1.7cm} c} \hline
\multicolumn{2}{c}{} & \multicolumn{7}{c}{Haar} & \multicolumn{7}{c}{Coiflet4}  \\
\cline{3-16}
\multicolumn{2}{c}{} &&diagonal && horizontal && vertical &&& diagonal && horizontal && vertical & \\ \hline
\multirow{2}{*}{OLS} & $H$ && \underline{\textbf{0.446}} && \underline{\textbf{0.481}} && \underline{\textbf{0.481}} &&& \underline{\textbf{0.488}} && \underline{\textbf{0.484}} && \underline{\textbf{0.481}} &\\
& \textit{MSE} && \cellcolor{black!20}0.004 && 0.002 &&  0.002 &&& \cellcolor{black!20}0.001 && \cellcolor{black!20}0.001 && \cellcolor{black!20}0.001 & \\ \hline
\multirow{2}{*}{AV} & $H$ &&  \textbf{0.385} && \textbf{0.467} && \textbf{0.466} &&& \textbf{0.439} && \textbf{0.384} && \textbf{0.388} & \\
& \textit{MSE} && 0.014 && 0.002 && 0.002 &&& 0.004 && 0.015 && 0.015 & \\ \hline
\multirow{2}{*}{TT} & $H$ && \textbf{0.404} && \textbf{0.473} && \textbf{0.472} &&& \textbf{0.453} && \textbf{0.395} && \textbf{0.397} & \\
& \textit{MSE} && 0.009 && \cellcolor{black!20}0.001 && \cellcolor{black!20}0.001 &&& 0.002 && 0.013 && 0.012 & \\ \hline
\multicolumn{16}{c}{} \\  \hline
\multicolumn{2}{c}{} & \multicolumn{7}{c}{Daub6} & \multicolumn{7}{c}{Symmlet8}  \\
\cline{3-16}
\multicolumn{2}{c}{} &&diagonal && horizontal && vertical &&& diagonal && horizontal && vertical & \\ \hline
\multirow{2}{*}{OLS} & $H$ && \underline{\textbf{0.484}} && \underline{\textbf{0.473}} && \underline{\textbf{0.474}} &&& \underline{\textbf{0.488}} && \underline{\textbf{0.485}} && \underline{\textbf{0.482}} & \\
& \textit{MSE} && \cellcolor{black!20}0.002 && 0.002 && \cellcolor{black!20}0.002 &&& \cellcolor{black!20}0.001 && \cellcolor{black!20}0.001 && \cellcolor{black!20}0.001 & \\ \hline
\multirow{2}{*}{AV} & $H$ && \textbf{0.436} && \textbf{0.456} && \textbf{0.456} &&& \textbf{0.440} && \textbf{0.401} && \textbf{0.401} & \\
& \textit{MSE} && 0.004 && 0.002 && 0.002 &&& 0.004 && 0.011 && 0.011 & \\ \hline
\multirow{2}{*}{TT} & $H$ && \textbf{0.451} && \textbf{0.463} && \textbf{0.463} &&& \textbf{0.454} && \textbf{0.409} && \textbf{0.409} & \\
& \textit{MSE} && 0.003 && \cellcolor{black!20}0.002 && 0.002 &&& 0.002 && 0.009 && 0.009 & \\ \hline
\end{tabular}}
\label{tab:2d_original}
\end{center}
\end{table}

\begin{table}[!h]
\begin{center}
\caption{Estimations of $H$ and MSEs for $H=0.5$ from three directions; under four different wavelet filters, in the contaminated case.}
\scalebox{0.9}{\begin{tabular}{c  l  c>{\centering}p{1.7cm}c>{\centering}p{1.7cm}c>{\centering}p{1.7cm}c  c>{\centering}p{1.7cm}c>{\centering}p{1.7cm}c>{\centering}p{1.7cm} c} \hline
\multicolumn{2}{c}{} & \multicolumn{7}{c}{Haar} & \multicolumn{7}{c}{Coiflet4}  \\
\cline{3-16}
\multicolumn{2}{c}{} &&diagonal && horizontal && vertical &&& diagonal && horizontal && vertical & \\ \hline
\multirow{2}{*}{OLS} & $H$ && \underline{\textbf{0.549}} && \textbf{0.571} && \textbf{0.568} &&& \textbf{0.584} && \textbf{0.583} && \textbf{0.588} & \\
& \textit{MSE} && \cellcolor{black!20} 0.004 && 0.007 && 0.007 &&& 0.008 && 0.009 && 0.009 & \\ \hline
\multirow{2}{*}{AV} & $H$ && \textbf{0.399} && \textbf{0.476} && \textbf{0.477} &&& \textbf{0.455} && \textbf{0.403} && \textbf{0.394} & \\
& \textit{MSE} && 0.011 && 0.001 && 0.001 &&& 0.002 && 0.011 && 0.013 & \\ \hline
\multirow{2}{*}{TT} & $H$ && \textbf{0.429} && \underline{\textbf{0.490}} && \underline{\textbf{0.493}} &&& \underline{\textbf{0.480}} && \underline{\textbf{0.423}} && \underline{\textbf{0.415}} & \\
& \textit{MSE} && 0.005 && \cellcolor{black!20} 0.001 && \cellcolor{black!20}0.001 &&& \cellcolor{black!20}0.001 && \cellcolor{black!20}0.008 && \cellcolor{black!20}0.009 & \\ \hline
\multicolumn{16}{c}{} \\  \hline
\multicolumn{2}{c}{} & \multicolumn{7}{c}{Daub6} & \multicolumn{7}{c}{Symmlet8}  \\
\cline{3-16}
\multicolumn{2}{c}{} &&diagonal && horizontal && vertical &&& diagonal && horizontal && vertical & \\ \hline
\multirow{2}{*}{OLS} & $H$ &&  \textbf{0.591} && \textbf{0.571} && \textbf{0.572} &&& \textbf{0.586} && \textbf{0.583} && \textbf{0.585} & \\
& \textit{MSE} && 0.010 && 0.006 && 0.007 &&& 0.009 && 0.008 && 0.009 & \\ \hline
\multirow{2}{*}{AV} & $H$ &&  \textbf{0.447} && \textbf{0.470} && \textbf{0.468} &&& \textbf{0.452} && \textbf{0.412} && \textbf{0.416} & \\
& \textit{MSE} && 0.003 && 0.001 && 0.001 &&& 0.003 && 0.009 && 0.008 & \\ \hline
\multirow{2}{*}{TT} & $H$ && \underline{\textbf{0.473}} && \underline{\textbf{0.487}} && \underline{\textbf{0.486}} &&& \underline{\textbf{0.476}} && \underline{\textbf{0.430}} && \underline{\textbf{0.434}} & \\
& \textit{MSE} && \cellcolor{black!20}0.001 && \cellcolor{black!20}0.000 && \cellcolor{black!20}0.001 &&& \cellcolor{black!20}0.001 && \cellcolor{black!20}0.006 && \cellcolor{black!20}0.006 & \\ \hline
\end{tabular}}
\label{tab:2d_contaminated}
\end{center}
\end{table}

Note that simulation results of fBms generated with $H=0.5$ are presented. Results for $H \in \{0.3, 0.4, 0.6, 0.7\}$ are not provided here because of space considerations; the simulation results for these values of $H$ can be found
at Jackets's Wavelet Page: \url{http://gtwavelet.bme.gatech.edu/datasoft/AppendixB.pdf}. The results are consistent when $H < 0.5$, where TT consistently outperforms OLS or AV in both 1-D and 2-D cases. However, for $H > 0.5$, the results were mixed.

\vskip 4mm

\noindent 4. Theil-type Estimation of Scaling in Breast Cancer Diagnostics

The scaling phenomenon has been found in many types of medical imaging and extensive research has been done utilizing this scaling for diagnostic purposes. Numerous references can be
found at \url{http://www.visionbib.com/bibliography/medical857.html}.

 Despite an overall reduction in the number of breast cancer cases, breast cancer still continues to be a major health concern among women. The National Cancer Institute estimates that 1 in 8 women born today will be diagnosed with breast cancer during her lifetime \citep{NCISEER2010}. One of the most important challenges is the increase of precision of screening technologies, since early detection remains the best strategy for improving prognosis and also leads to less invasive options for both specific diagnosis and treatment.

\vskip 4mm

\noindent 4.1 Description of the Data Set

\begin{figure}[!htb]
\centering
\includegraphics[width=2.4in,height=2.4in]{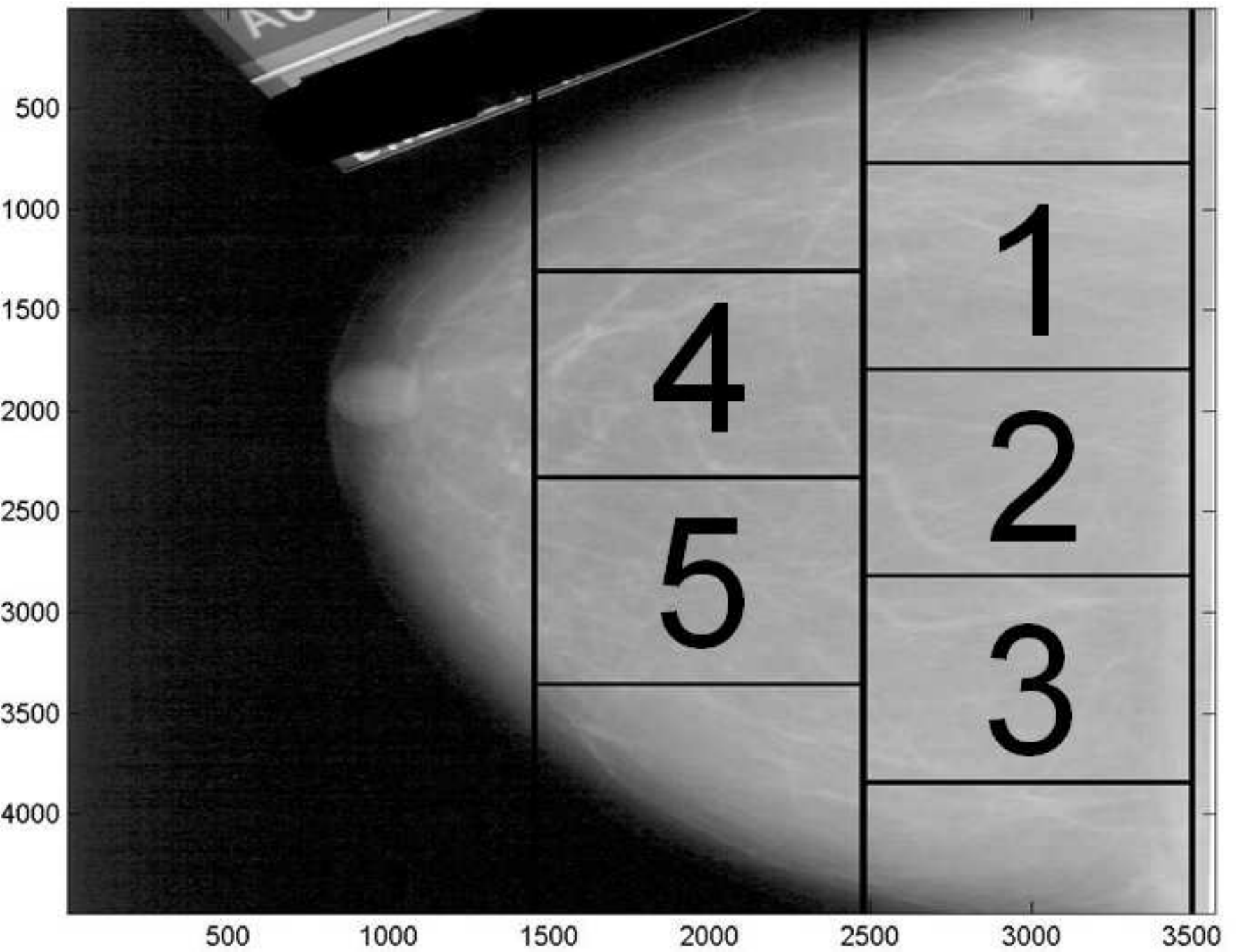}
\caption{ Five subimages of size $1024 \times 1024$ are extracted from each breast image to capture tissues from the designated locations.}
\label{fig:mammo}
\end{figure}

A collection of digitized mammograms for analysis was obtained from the University of South Florida's Digital Database for Screening Mammography (DDSM). The DDSM is described in detail in \cite{Heath2000}. Images from this database containing suspicious areas are accompanied by pixel-level ``ground truth'' information relating locations of suspicious regions to what was assessed and verified through biopsy.  45 normal cases (controls) and 79 cancer cases scanned on the HOWTEK scanner at the full 43.5 micron per pixel spatial resolution were selected. Each case contains four mammograms from a screening exam, two projections for each breast: the craniocaudal (CC) and mediolateral oblique (MLO).  Only the CC projections were considered, using either side of the breast image. Five subimages of size 1024 $\times$ 1024 were taken from the mammograms. An example of a breast image and location of subimages is provided in Fig. \ref{fig:mammo}. Black lines that compart the breast area into 5 squares in a mammogram show how subimages are sampled from the original image.

\vskip 4mm

\noindent 4.2 Estimation of $H$

For every subimage, the DWT using Haar and Symmlet 8-tap filters were performed to observe sensitivity in results under different wavelet bases. The analysis was repeated with four different sets of levels used for the regresion: 4 to 9, 5 to 9, 6 to 9, and 7 to 9. After each transform, OLS, AV, and TT estimation methods were used to compute the directional Hurst exponents, $H_d$, $H_h$, and $H_v$.

 A nested  ANOVA  was performed to test if the Hurst estimates have significant differences based on the health condition of a patient,
\ba
H_{ijk} = \mu + \alpha_i +\beta_{j(i)}+\epsilon_{ijk},
\ea
where $i$ indicates the health status of a patient ($i$=1 for cancer, $i$=2 for normal), $j(i)$ indicates a patient  nested in the status $i$, ($j(1)  = 1,  \dots, 45; j(2) = 1,\dots, 79$), and $\epsilon_{ijk}$ is an error term ($k=1,\dots, 5$). Table \ref{tab:anova} summarizes the results of the ANOVA analysis on diagonal Hurst exponents ($H_d$) obtained using Symmlet 8 tap filter and the TT method.

\begin{table}[!h]
\begin{center}
\caption{ANOVA results on $H_d$ using Symmlet 8 and the TT method.}
\vspace{.1in}
\begin{tabular}{l c c c c c} \hline
Source & Sum Sq. & d.f. & Mean Sq. & F & p-value \\  \hline
Status &   0.330  &   1 &  0.330  & 10.741  &    0.001 \\
Patients(Status)   & 3.750  & 122  &  0.031 & 7.574 & $<$0.001 \\
Error   &   2.013    &   496   &   0.004 & &    \\  \hline
Total   &   6.093    &   619    & & &\\
\hline
\end{tabular}
\label{tab:anova}
\end{center}
\end{table}

Note that 5 images are taken for each subject. This gives a total number of images $(45+79)\times 5 = 620.$
However, since the subjects had multiple images and were used as blocks within the disease status,
the nested ANOVA was necessary to correctly analyze this data.

 The ANOVA model used was also demonstrated to be appropriate by checking for the normality and independence of residuals. The $n=620$ residuals at each hierarchy $H_d$, $H_h$ and $H_v$ conformed to a battery of standard goodness-of-fit and independence tests. We were particulary focused on the deviations from the symmetry of residuals, given non-robustness of standard ANOVA to alternatives of asymmetry.  To this end, the Lin-Mukholkar and Jarque-Bera tests were conducted and found not significant. At first glance the independence is a non-issue here, since the observations could be freely permuted in each of the disease classes. This, however, is not the case since the positions of subimages 1-5 are comparable within each mammogram. The Durbin-Watson test of independence against the order of residuals was found insignificant as well.

The $p$-values for disease factor from the ANOVA analysis were 0.001 for $H_d$ (Table \ref{tab:anova}), 0.003 for $H_h$, and 0.020 for $H_v$. Based on the ANOVA analysis, we conclude that a patient's health condition is a significant factor that affects Hurst exponents in any of the three directions: diagonal, horizontal, and vertical. In the subsequent classification procedure based on ANOVA estimators of disease status, it is shown that these main effects are not only statistically significant, but also discriminatory.

\vskip 4mm

\noindent 4.3 Classification  {of Images}

To classify mammogram images as cancerous or non-cancerous, the estimated $H_d$ and pair ($H_d, H_h$) for each subject were taken using nested ANOVA. Operationally, this is $H_{ij, d} = \mu + \alpha_i +\beta_{j(i)}.$

Next, the subjects were classified by disease status using a logistic regression with four fold cross validation. The classification was repeated 300 times and the results were averaged over these 300  repetitions. A threshold of the logistic regression based on the maximum Youden index was chosen, which indicates the threshold (i.e., 0.6057) providing the maximum true positive and true negative accuracy.

 \begin{table}[!htb]
 \centering
 \caption{Results of classification by logistic regression using $H_d$ and $H_d, H_h$ }
\vspace{.1in}
 \begin{tabular}{l c ccc c ccc}
 \hline
 Predictors & & \multicolumn{3}{c}{$H_d$} & & \multicolumn{3}{c}{$H_d$, $H_h$}\\ \hline
	 \small Method & &\small Total ~ &\small Specificity ~ &\small  Sensitivity ~  & &\small Total~ &\small Specificity ~ &\small Sensitivity ~  \\ \hline
 \small OLS  & &\small 0.648  &\small0.567  &\small  0.682 & &\small 0.626 &\small 0.554  &\small 0.707 \\
 \small AV   & &\small 0.652 &\small 0.511 &\small  0.733 & &\small 0.640 &\small 	 0.502 &\small  0.747  \\
 \small TT   & &\small 0.654   &\small  0.550  &\small 0.709 & &\small 0.639 &\small  0.537  &\small  0.725  \\
 \hline

 \end{tabular}
 \label{tab:auc}
 \end{table}

Table \ref{tab:auc} summarizes the results of the classification based on $H_d$ and ($H_d, \;H_h$), for each estimation method. The first column provides total classification accuracy, while the next two columns provide true positive and true negative rates. The best classification rates were achieved with the TT and AV estimators, where the classification error was around 35\% for both methods. OLS was the worst performer, with 36\% error.

Unlike the simulation cases where the monofractality of the signals were violated by design, and where
the TT method was clearly favored,
for mammograms the AV and TT methods performed comparably. This may be the consequence of the fact that
individual $1024 \times 1024$ subimages taken from the mammograms exhibited reasonable isotropy and
monofractality, with wavelet coefficients being aproximately Gaussian.

\begin{figure}[!htb]
\begin{center}
 \includegraphics[width=3in,height=2.8in]{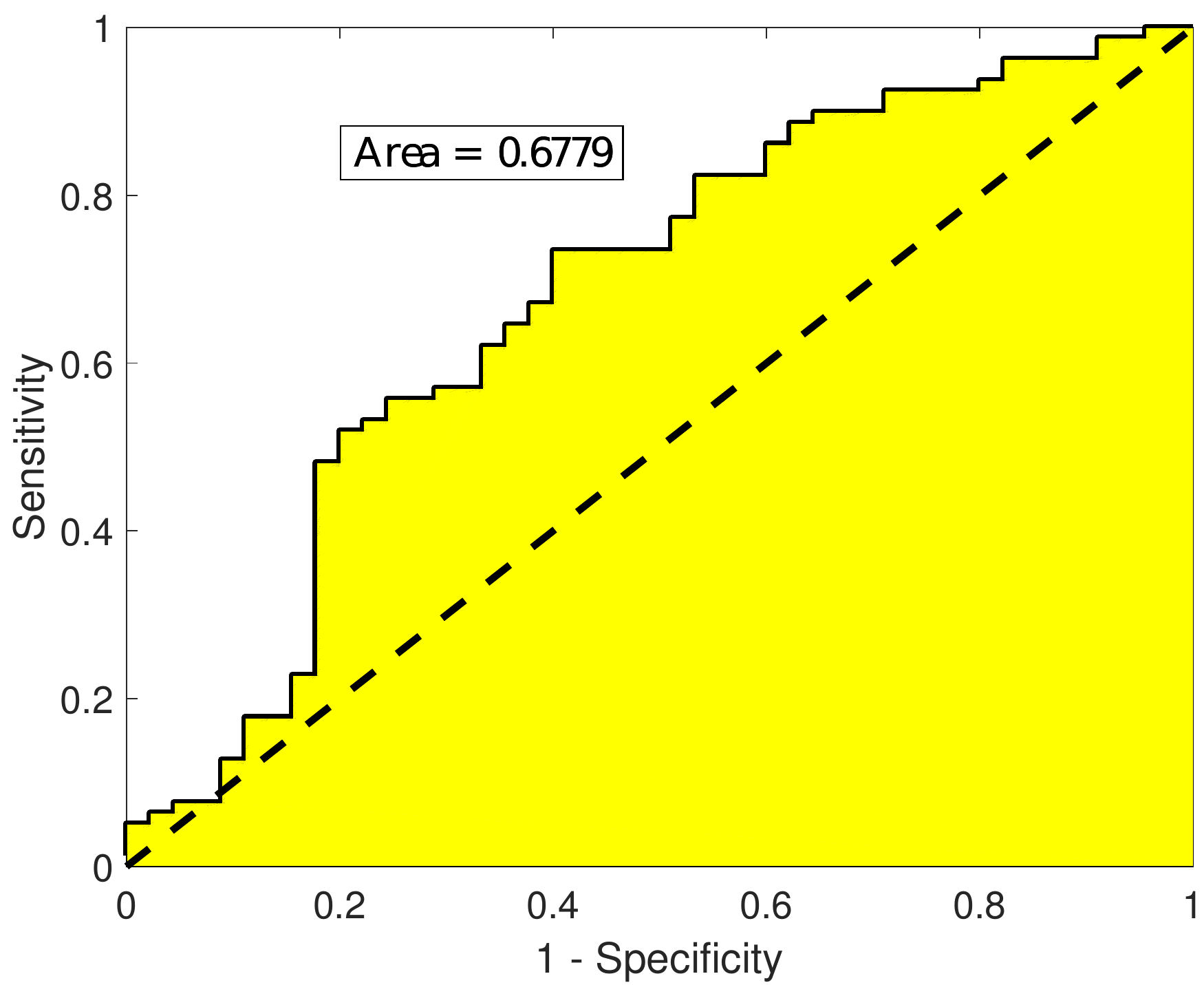}
\caption{ROC curve for the logistic regression:$ \mbox{logit}(p) =     5.027 -7.968 \cdot H_d $. }
\label{fig:ROC}
\end{center}
\end{figure}

Figure \ref{fig:ROC} shows a ROC curve of $H_d$ (obtained by TT) in differentiating between controls and cancer cases. The diagonal line represents a test with a sensitivity of 50\% and a specificity of 50\%. This shows the ROC curve lying significantly to the left of the diagonal, where the combination of sensitivity and specificity are highest. The area under the ROC curve, which is proportional to the diagnostic accuracy of the test, is 0.678.

\vskip 4mm

\noindent 5. Discussion \& Conclusions

In this paper a novel wavelet-based
Theil-type (TT) robust estimator of scaling was presented, with theoretically optimal weights for pairwise slopes
that depend on harmonic average of sample sizes from the two multiresolution levels defining the pair.
This estimator is free of distributional assumptions for the underlying regression, and robust with
respect to possible outlier levels.
An extensive simulational study demonstrated that the TT estimator was comparable, and in many scenarios
superior, to the standardly used ordinary least squares (OLS) and Abry-Veitch (AV) estimators.
This superiority is reflected in both less bias and smaller mean-squared error.

In the context of mammogram classification, the TT method was found to be comparable to AV and superior to OLS methods.
This closeness of TT and AV could be explained by apparent relative spatial homogeneity and monofractality of the mammogram subimages used in the analysis.

For all three estimators considered, adding spectral indices from directional hierarchies other than the diagonal did not always improve the diagnostic performance. Index $H_d$ by itself was strongly discriminatory and the most parsimonious classifying summary. Furthermore, the results of classification using all three spectra ($H_d, H_h$, and $H_v$) did not always perform better, on average, than that with only one or two spectral indices.

The diagnostic use of information contained in the background of images is often an ignored modality.
It  allows for the use of information from the entire image, rather than focusing primarily on irregular shapes, masses, or calcifications.


       Recently, many studies have proposed fractal based modeling to describe and detect the pathological architecture of tumors.
        For example, the authors in \cite{Hermann2015} demonstrated breast cancer screening using fractal and stochastic geometric approaches such as random carpets, Quermass-interaction process, and complex-wavelet based self-similarity measures.

       Although medical images exhibit high heterogeneity attributing to their multifractality, the use of the robust estimator proposed in this paper has proven that the monofractal self-similarity measure can be a promising classifier to differentiate malignant images from benign. As it is important to combine several instruments for cancer testing, this paper provides a quick and robust quantitative measure to strengthen existing mammogram classification procedures. Although the accuracy rates could be argued to be relatively low, even classifiers that are ``slightly better than flipping a coin'' can improve diagnostic accuracy when added to a battery of other independent testing modalities.

 Finally, as future work, the authors plan to extend \cite{Pepa2013}, where the multifractal spectrum was used for diagnostic classification in a similar context. Some experiments in this direction were conducted  in
 \cite{BraniPepa2012}, however, direct comparisons with the results in this paper were not possible because of
 different data sets used.

\vskip 4mm

\noindent {\bf Acknowledgement.~} The corresponding author thanks the projects MTM2015-65915 (Ministerio de Econom\'ia y Competitividad, Spain), P11-FQM-7603 and FQM-329 (Junta de Andaluc\'ia), all with EU ERD Funds.
Brani Vidakovic was supported by NSF-DMS 1613258 award  and
 the National Center for Advancing Translational Sciences of the National Institutes of Health under award UL1TR000454.

\newpage

\noindent BIBLIOGRAPHY
\bibliographystyle{apalike}
\bibliography{Theil}

\newpage

\noindent APPENDIX A: Derivation of the weights of the TT approach

Let $d_j = d_{j{\mathbf{k}}}$ be an arbitrary (wrt $\mathbf{k}$) wavelet coefficient from
the $j$th level of the decomposition of the $m$-dimensional fractional Brownian motion $B_H(\omega, \mathbf{t}), \mathbf t \in \R^m$,
\ba
d_j = \int_{\R^m}~ B_H(\omega, \mathbf{t}) \psi^*_{j\mathbf{k}}(\mathbf{t}) d\mathbf{t},
\ea
for some fixed $\mathbf{k}=(k_1, \dots, k_m).$
Here $\psi^*_{j\mathbf{k}}(\mathbf{t}) = \prod_{i=1}^m \psi^*_{jk_i}(t_i)$ where $\psi^*$ is
either $\psi$ or $\phi$, but in the product there is at least one $\psi.$
It is well known that
\ba
d_j \stackrel{d}{=} 2^{-(H + m/2) j } ~d_0,
\ea
where $d_0$ is a coefficient from the level $j=0,$ and $\stackrel{d}{=}$
means equality in distributions.

Coefficient $d_j$ is a random variable with
\ba
\EE d_j = 0 \; ~\text{and} \;~ \var d_j = \EE d_j^2 =  2^{-(2 H + m) j } ~\sigma^2,
\ea
where $\sigma^2 =   \var d_0.$

The fBm $B_H(\omega, \mathbf t)$ is a Gaussian $m$-dimensional field, thus
\ba
d_j \sim {\cal N}(0, 2^{-(2 H + m )j} \sigma^2 ).
\ea
The coefficients $d_j$ within the level $j$ are typically considered
approximately independent. The covariance decays with the distance
between the coefficients and the rate of decay depends on
$H$ and $N$ - the number of vanishing moments for the wavelet $\psi.$
 \cite{Flandrin1992} and \cite{Tewfik1992} showed that for $m=1$,
\ba
\EE d_{jk_1} d_{jk_2} \leq C |k_1 - k_2|^{2(H-N)},
\ea
where $C$ depends on $j.$
Although, for small $|k_1 - k_2|$ this covariance may not be small,
it decays to 0 as long as $N > H.$
To ensure short memory of $d_{jk}, ~k \in \ZZ,$ the convergence of
\ba
\sum_{k} \EE |d_{jk_1} d_{jk_2}|
\ea
is needed, for which it is required that $N > H + 1/2.$

The rescaled  ``energy''
\ba
 \frac{2^{(2 H + m )j} }{\sigma^2}  d_j^2 \sim \chi^2_1 ,
 \ea
 while, assuming the independence of $d_{jk}$'s,
 \ba
  \frac{ 2^{(2 H + m )j} }{\sigma^2} \sum_{\mathbf k \in
  \mbox{\footnotesize $j$th level}} d_{j\mathbf k}^2
  =  \frac{2^{(2 H + 2 m)j} }{\sigma^2} ~ \overline{d_{j}^2} ,
 \ea
 has $ \chi^2_{2^{mj}}$ distribution.
Here, $ \overline{d_{j}^2}$ is the average energy in $j$th level.

Thus,
\ba
 \overline{d_{j}^2} \stackrel{d}{=}  2^{-(2 H + 2 m )j } \sigma^2 \chi^2_{2^{m j}}  .
\ea

From this,

\ba
\EE \overline{d_{j}^2} = \sigma^2 2^{-(2 H + 2 m)j } \EE \chi^2_{2^{m j}} =
 2^{-(2 H + m )j } \sigma^2  ,
 \ea
and
\ba
\var \overline{d_{j}^2} = \sigma^4 2^{-(4 H + 4 m)j } \times 2 \cdot 2^{m j} =
 2^{-4 H j - 3 m j + 1} \sigma^4 .
\ea

Recall that if $X$ has $\EE X$ and $\var X$ finite and $\varphi$ is
a function with finite second derivative at $\EE X$,
then
\ba
\EE\varphi(X) \approx  \varphi(\EE X) + \frac{1}{2} \varphi''(\EE X) \cdot \var X ,
\ea
and
\ba
\var \varphi(X) \approx (\varphi'(\EE X))^2 \var X.
\ea

When $\varphi$ is logarithm for base 2, then
\ba
\EE \log_2 \overline{d_{j}^2} & = & \log_2 \EE \overline{d_{j}^2} +
\frac{1}{2 \log 2} \left( - \frac{\var \overline{d_{j}^2} }{\left(\EE \overline{d_{j}^2}\right)^2} \right) \\
& = &\log_2\left(2^{-(2 H + m )j } \sigma^2 \right)  - \frac{1}{2 \log 2} 2^{-m j + 1}\\
& = &-(2 H + m )j - \frac{1}{2^{mj} \log 2}  + \log_2 \sigma^2.
\ea
Note that  $- \frac{1}{2^{mj} \log 2}$ is the
Abry-Veitch bias term, and it is
free of $H$ and $\sigma^2.$ This bias is a second order approximation.
Veitch and Abry show that the exact bias involves digamma function
$\Psi$, and in this context is
\ba
\frac{\Psi(2^{mj-1})}{\log 2} - \log\left(2^{mj-1}\right).
\ea

Also,
\ba
\var \log_2 \overline{d_{j}^2}&=&\left( \frac{1}{\sigma^2~ 2^{-(2 H + m )j } ~\log 2  }\right)^2\\
& & ~~~\times \sigma^4 \cdot 2^{-4Hj - 3mj + 1}  \\
&=&\frac{2}{2^{mj} (\log 2)^2}.
\ea

Finally,
\ba
\var\left( \frac{\log_2 \overline{d_{j}^2} - \log_2 \overline{d_{i}^2}}{j - i }\right)
= \frac{2}{(\log 2)^2} \cdot \frac{1/2^{mj} + 1/2^{mi}}{(j-i)^2}.
\ea
Since weights $w_{ij}$ are inverse-proportional
to the variance, then
\ba
w_{ij} \propto (i-j)^2 \times HA(2^{mi}, 2^{mj}),
\ea
where $HA$ is the harmonic average.

\end{document}